\documentclass{article}
\usepackage[utf8]{inputenc} % allow utf-8 input
\usepackage[T1]{fontenc}    % use 8-bit T1 fonts
\usepackage{hyperref}       % hyperlinks
\usepackage{url}            % simple URL typesetting
\usepackage{booktabs}       % professional-quality tables
\usepackage{amsfonts}       % blackboard math symbols
\usepackage{nicefrac}       % compact symbols for 1/2, etc.
\usepackage{microtype}      % microtypography

\usepackage{authblk}
\usepackage{natbib}
\usepackage[margin=1in]{geometry}
\usepackage[utf8]{inputenc} % allow utf-8 input
\usepackage[T1]{fontenc}    % use 8-bit T1 fonts
\usepackage{hyperref} 

\usepackage{url}            % simple URL typesetting
\usepackage{booktabs}       % professional-quality tables
\usepackage{amsfonts}       % blackboard math symbols
\usepackage{nicefrac}       % compact symbols for 1/2, etc.
\usepackage{microtype}      % microtypography
\usepackage{xcolor}         % colors
\title{Incentivized Collaboration in Active Learning}

\usepackage[utf8]{inputenc} % allow utf-8 input
\usepackage[T1]{fontenc}    % use 8-bit T1 fonts
\usepackage{hyperref}       % hyperlinks
\usepackage{url}            % simple URL typesetting
\usepackage{booktabs}       % professional-quality tables
\usepackage{amsfonts}       % blackboard math symbols
\usepackage{nicefrac}       % compact symbols for 1/2, etc.
\usepackage{microtype}      % microtypography
\usepackage{xcolor}         % colors
\usepackage{hyperref}
\usepackage{amsmath,amsfonts,amssymb}
\usepackage{bm}
\usepackage{wrapfig}

\usepackage{enumitem,array,booktabs}
\usepackage{algorithm}
\usepackage{algorithmic}
\usepackage{bbm}
\usepackage{tabulary,multirow}
\usepackage{dsfont}
\usepackage{natbib}
\usepackage{tikz}
\usepackage{mdframed}
\usepackage{thm-restate}

\DeclareMathOperator*{\argmax}{arg\,max}

\newcommand{\NN}{{\mathbb N}}

\newcommand{\ind}[1]{\mathds{1}[#1]}

\newcommand{\EEs}[2]{\mathbb{E}_{#1}\left[#2\right]}

\newcommand{\abs}[1]{\left|#1\right|}

\newcommand{\cA}{\mathcal{A}}
\newcommand{\cB}{\mathcal{B}}

\newcommand{\cF}{\mathcal{F}}

\newcommand{\cH}{\mathcal{H}}

\newcommand{\cO}{\mathcal{O}}

\newcommand{\bigO}{\mathcal{O}}

\newcommand{\cR}{\mathcal{R}}

\newcommand{\cT}{\mathcal{T}}

\newcommand{\cX}{\mathcal{X}}

\newcommand{\cY}{\mathcal{Y}}

\renewcommand{\epsilon}{\varepsilon}
\renewcommand{\hat}{\widehat}

\newcommand{\nothere}[1]{}

\newcommand{\VS}{\mathrm{VS}}

\newcommand{\opt}{\textrm{OPT}}
\newcommand{\OPT}{\textrm{OPT}}

\newcommand{\gbs}{\textrm{GBS}}

\newcommand{\Xset}{X_1,\ldots,X_k}
\newcommand{\xset}{\{\Xset\}}

\newcommand{\ir}{\textrm{B2IR}}

\author[]{Lee Cohen}
\author[]{Han Shao}
\affil[]{Toyota Technological Institute of Chicago\\
{ \texttt{\{lee, han\}@ttic.edu}}}
\date{}

\usepackage{amsthm}
\newtheorem{definition}{Definition}
\newtheorem{lemma}{Lemma}
\newtheorem{theorem}{Theorem}
\newtheorem{corollary}{Corollary}

\newtheorem{assumption}{Assumption}
\newtheorem{example}{Example}[section]

\begin{document}

\maketitle

\begin{abstract}
% \hs{TODO:
% \\
% 1. add a high-level sketch of the example in the proof of ``Greedy is not IR''. maybe also describe why OPT is IR in the same example
% \\
% 2. reduce the length of converting alg to IR/SIR. 
% \\
% 3. add a tree to illustrate the algorithmic idea, e.g., how to truncate the tree
% }

In collaborative active learning, where multiple agents try to learn labels from a common hypothesis, we introduce an innovative framework for incentivized collaboration. Here, rational agents aim to obtain labels for their data sets while keeping label complexity at a minimum. We focus on designing (strict) \emph{individually rational} (IR) collaboration protocols, ensuring that agents cannot reduce their expected label complexity by acting individually. We first show that given any optimal active learning algorithm, the collaboration protocol that runs the algorithm as is over the entire data is already IR. However, computing the optimal algorithm is NP-hard. We therefore provide collaboration protocols that achieve (strict) IR and are comparable with the best known tractable approximation algorithm in terms of label complexity.

\end{abstract}

% \begin{keywords}%
%    individual rationality, incentives, pool-based active learning, Bayesian learning, collaboration
% \end{keywords}

\section{Introduction}\label{sec:intro}
Active learning has emerged as a powerful paradigm in which labels of selected data points are sequentially queried from a large pool of unlabeled data, referred to as the unlabeled pool. The primary objective is to minimize labeling effort to find a classifier that exhibits low error on fresh data points from the same data source, known as generalization error. 
Typically, if the pool is large enough, a classifier that performs well on the pool can also achieve low generalization error through uniform convergence. 

Active learning has also been studied in the distributed setting, where the unlabeled pool is scattered across multiple machines (called agents), (e.g.,~\cite{shen2016distributed,aussel2020combining}). 
While active learning has demonstrated promising results, traditional approaches often operate in isolation, neglecting the potential benefits 
of collaboration among agents should they agree to collaborate. In this paper, we propose a novel framework for incentivized collaboration active learning, where agents can collaboratively explore their data pools to discover a common target function.

The motivation for collaboration in active learning stems from real-life scenarios where collaboration and collective intelligence yield improved outcomes, e.g., when agents collect data from the same distribution, and can easily end up labeling the same or very similar points. This redundancy leads to unnecessary and inefficient utilization of resources, as the labeling is often done by experts. Additionally, more data can be translated to improved accuracy, prompting agents to pool their resources and employ a more powerful model.

The incentive-driven nature of our framework aligns with the reality of collaboration in the real world. When agents are incentivized to collaborate only when their expected labeling complexity decreases, it reflects the real-life scenario where individuals are motivated to engage in cooperative endeavors if they perceive a clear benefit, such as reduced effort, faster, and better outcomes. In this work, we focus on a specific notion of incentives, where agents already have access to a baseline algorithm and they are motivated to join the collaboration if their label complexity is smaller than running the baseline algorithm on their own.

Consider, for example, the case of a new drug (e.g., Paxlovid for Covid-19~\cite{Paxlovid22}, that has different efficacy on patients with different features. While individual hospitals can test the drug on their patients in an active learning fashion by executing their preferred baseline algorithm, collaborating efficiently with other hospitals, each with their own patients, often leads to a better prognosis. 

However, if the incentives of the hospitals are not maintained, i.e., the effort of some hospitals is increased, the collaboration may be compromised. By emulating this collaboration within the active learning framework, we unlock the potential of collective intelligence to enhance the learning process. %\lc{We remark that in the context of medical applications, there could be other constraints (e.g., age, medical condition) that play a role in the choice of who should receive the drug, which we do not consider in this work}
Besides, imagine that several data labeling companies have to recover the labels of unlabeled images assigned to them. Each data labeling company would like to collaborate with other companies to recover the labels of all images while minimizing the query complexity and not increasing their burden.

Our basic model is as follows: there are $k$ agents, each with their own set of unlabeled data points, and a single hypothesis class with a prior on the hypotheses, which all agents are aware of.
We assume realizability, meaning that there exists an underlying ground truth labeling function called the target function, labeling all the data points, and that the hypothesis class encompasses such a target function. We refer readers to the discussion for more information about this assumption in the context of active learning.

The agents reach a consensus on an arbitrary baseline algorithm for pool-based active learning (e.g., the best tractable approximation algorithm).
To select whether or not to join the collaboration, the agents need to evaluate their utility from joining the collaboration. Since the goal of each individual agent (regardless of the collaboration) is to minimize their expected query complexity, the most natural cost function is the expected query complexity.
To ensure each individual benefits from their collaboration, we establish a collaboration protocol that guarantees that each agent cannot reduce their expected label complexity by running the baseline algorithm individually. This concept is referred to as individual rationality (IR).
% , \lc{and when it is satisfied, the collaboration protocol is Pareto optimal.}.
Our objective is to design an IR collaboration protocol that minimizes the overall labeling queries.

% [interesting examples]

There are cases in which collaboration is not necessarily beneficial. For example, each agent has non-zero points on a different axis and the hypothesis class contains every possible halfspace. If the prior distribution is uniform over all labelings, then no agent can reduce their label complexity by joining the collaboration.

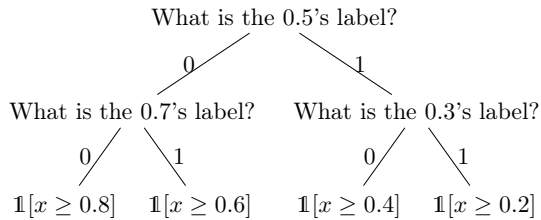
\begin{wrapfigure}[12]{R}{0.5\textwidth}
% \vspace{-2mm}
    \begin{minipage}{0.5\textwidth}
% \begin{figure}[t]
    \centering
    \scalebox{0.9}{
\begin{tikzpicture}[level distance=1.4cm,
  level 1/.style={sibling distance=4.2cm},
  level 2/.style={sibling distance=2cm}]

  \node (root) {What is the $0.5$'s label?}
    child {node (A) {What is the $0.7$'s label?}
      child {node (B) {$\ind{x\geq 0.8}$}}
      child {node (C) {$\ind{x\geq 0.6}$}}}
    child {node (D) {What is the $0.3$'s label?}
      child {node (E) {$\ind{x\geq 0.4}$}}
      child {node (F) {$\ind{x\geq 0.2}$}}};

  % Labels
  \path (root) -- (A) node[midway,left] {$0$};
  \path (root) -- (D) node[midway,right] {$1$};
  \path (A) -- (B) node[midway,left] {$0$};
  \path (A) -- (C) node[midway,right] {$1$};
  \path (D) -- (E) node[midway,left] {$0$};
  \path (D) -- (F) node[midway,right] {$1$};
\end{tikzpicture}}
    \caption{The query tree of binary search for thresholds.}
    \label{fig:query-tree}
% \end{figure}
\end{minipage}
\end{wrapfigure}
Clearly, if each agent has the same set of points, the label complexity of each agent can decrease to $1/k$ of its original label complexity if the collaboration protocol equally splits the labeling burden.
Even if agents do not share the same set of points, they can still benefit from collaboration, as we show in the next example.
% \lc{[how about we put it in an example environment and refer to it after assumption 2?]} 
\begin{example}\label{ex:treshold}
Consider the scenario with 1-dimensional thresholds $\cH =\{\ind{x\geq \alpha}|\alpha = 0.2,0.4,0.6,0.8\}$ and a uniform prior distribution over $\cH$. Suppose agent 1 has points $\{0.25,0.5,0.75\}$ and agent 2 has points $\{0.3,0.45, 0.55, 0.7\}$. 
When running binary search collaboratively, each agent only performs one labeling query, as illustrated as a search tree in Fig~\ref{fig:query-tree}. On the other hand, if they were to run binary search independently, each agent would need to query $2$ labels. Thus, collaboration can effectively reduce the label complexity for each agent by $1$.
\end{example}

 \textbf{Bayesian Assumption.} 
 The reason why we have a Bayesian assumption regarding the hypothesis class is that without it, querying all the labels to discover the target hypothesis can be inevitable, even for a simple class of linear separators in $\cR^2$ (see, e.g., Claim $1$ in~\cite{dasgupta2004analysis}).
It is worth noting that as in~\cite{dasgupta2004analysis}, we \textit{do not require the prior distribution to align with nature}. Instead, the prior distribution serves as a measure for average case analysis. Having a prior belief in our model has the following clear assumption. If the algorithm reaches a point where the remaining consistent hypotheses largely agree 
on the unlabeled data, it is reasonable to stop and output one of these remaining hypotheses~\cite{FreundSST97}. In a non-Bayesian setting, it does not make sense to operate this way. 

\textbf{Game Theory Interpretation.} 
%https://en.wikipedia.org/wiki/Mechanism_design
The agreed-upon baseline algorithm induces a sort of (not private) values for agents- each agent has its (negative) individual labeling complexity as value. 
The collaboration protocol can be then interpreted as a mechanism: 
Initially, the collaboration protocol (principal) is introduced to the agents, and each agent can understand it and have confidence in the principal's commitment to implementing it faithfully. Subsequently, the agents either rely on their trust in the algorithm's IR property or have the ability to verify it autonomously. Lastly, the agents behave rationally by joining the collaboration only if it is IR.

We remark that there is an interesting parallelism between our IR collaboration algorithms and truthful mechanisms. It is well known that Vickrey–Clarke–Groves (VCG) mechanism is a truthful mechanism that maximizes social welfare, but since it is hard to compute and to approximate~\cite{Buchfuhrer10}, the optimal outcome is replaced by a sub-optimal outcome of an approximation algorithm, and the resulting mechanism is not necessarily truthful. The goal is therefore relaxed to design an efficient approximation algorithm that returns a truthful mechanism.

\textbf{Contributions and Organization.} We formalize the model in Section~\ref{sec:model}. In Section~\ref{sec:ir}, we demonstrate that any optimal algorithm is individually rational when the baseline is itself. This implies that optimizing for optimality ensures individual rationality for all baseline algorithms.

However, computing (or even approximating) the optimal algorithm is known to be NP-hard. To address this, we then show that the best available tractable approximation algorithm, the greedy algorithm \cite{kosaraju2002optimal,dasgupta2004analysis}, is not individually rational when the baseline is itself.
We demonstrate this by presenting an example where joining the collaboration increases the labeling complexity of an agent from $O(1)$ to $\Omega(n)$. 
In response, we introduce a general approach that can transform any arbitrary baseline algorithm into an IR collaborative algorithm. This conversion ensures that the total label complexity remains competitive with running the baseline algorithm on the entire data set. Furthermore, in Section~\ref{sec:sir} we present a scheme that converts any IR collaborative algorithm into a strict IR one, guaranteeing the label complexity is strictly lower by joining the collaboration under mild assumptions. 
When the baseline algorithm is both efficient and approximately optimal, our (strict) IR algorithms efficiently achieve label complexity that is approximately optimal.

\paragraph{Related Work.}
% strategic behavior
The most related work is the recent work of \cite{xu2023fair}, which studies individual rationality in collaborative active learning with Gaussian Process models. While their notion of individual rationality is similar to ours, we focus on query complexity in binary classification. \cite{echenique2019incentive} study incentive compatibility in active learning in a setting with a single strategic agent who responds to the learner's queries. Another closely related work is \cite{balcan2014active}, which studies active learning in a multi agent setting where noisy distributed sensors use local communication and best response dynamics to denoise their labels. Their focus is on how game theoretic dynamics can reduce label noise and enable query efficient learning, whereas our work studies rational agents with individual data pools and designs collaboration protocols that satisfy individual rationality with respect to each agent's own label complexity. Our work is situated at the junction of learning in the presence of strategic behavior and active learning.

\paragraph{Learning in the presence of strategic behavior}
Learning in the presence of strategic behavior encompasses a vast body of research, including~\cite{ben-porat23,Zhang22,hardt2016strategic}.
We are particularly driven by prior research in this area, and by the question of how to create learning algorithms that incentivize agents to participate while maximizing the overall welfare. For example, \textit{incentivized exploration} has been studied in multi armed bandits~\cite{Kremer-JPE14,MansourSS15,MansourSSW16,MansourSW18,Cohen19,Che-13,BaharST16,Bahar2019FiduciaryB,Bahar19,Immorlica19,Immorlica20,Sellke21,banihashem2023bandit,Slivkins17,slivkins2019introduction} and MDPs~\cite{simchowitz2023exploration}, where the principal recommends actions to the agents in order to explore different alternatives, but the agents ultimately decide whether to follow the recommendation. This raises incentive constraints in addition to the exploration exploitation tradeoff. In particular, \cite{Baek21} study this problem in the context of fairness with a group based regret notion. They show that regret optimal bandit algorithms can be unfair and design a nearly optimal fair algorithm. Incentivizing agents to share their data has been studied by \cite{wei2023incentivized} in federated bandits.

\paragraph{Federated learning}
Federated learning has gained popularity as a method to foster collaboration among large populations of learning agents, among other reasons for incentivizing participation and fairness purposes~\cite{BlumHPS21,Lyu2020,Donahue_Kleinberg_2021,Donahue22Fair,DonahueK21,Donahue23,wang2023framework}. A related recent work is \cite{huang2023evaluating}, who study incentives for diverse data contributions in collaborative learning. Their work focuses on a federated learning setting where the goal is to incentivize agents to contribute data that is useful for learning a global model. In contrast, our work focuses on collaboration protocols for active learning and gives individual rationality guarantees with respect to the agents' label complexity. Our work also addresses fairness, in the sense that if a collaborative algorithm is individually rational, it is fair for all participating agents. Another related line of research is \textit{kidney exchange}~\cite{Roth04,AshlagiR11,BlumIHPPV17kidney,BlumG21,BlumMansour20,DickersonPS19}, where the goal is to find a maximum matching in a directed graph representing transplant compatibilities between patient donor pairs. In this problem, incentives arise in the form of individual rationality when different hospitals have different subsets of patient donor pairs and will not join the collaboration if the number of pairs matched through collaboration is lower than the number of pairs they could match on their own.

\paragraph{Active learning}
Active learning has a long history in learning theory. Foundational work includes agnostic active learning~\cite{balcan2006agnostic}; see also \cite{balcan2015active,Hanneke14survey} for surveys. There are two basic models in active learning: stream based active learning~\cite{FreundSST97}, where the learner must decide immediately whether to query the label of the current instance or discard it, and pool based active learning, which is the basis for our model. Pool based active learning studies settings in which a learner is given a collection of unlabeled data points and aims to recover a target function by adaptively querying labels. Active learning has also been studied in the context of other desiderata such as fairness~\cite{ShenCW22,Camilleri23} and safety~\cite{CamilleriWMJJ22}.

To our knowledge, no prior work has combined these directions to study individual rationality constraints in collaborative active learning with respect to label complexity. This is where our work makes a valuable contribution.

\section{ Preliminaries and Model}\label{sec:model}
Throughout the work, we consider the binary classification problem. 
Let $\cX$ denote the input space, $\cY =\{0,1\}$ denote the label space, and $\cH\subset \cY^\cX$ denote the hypothesis class.\footnote{Results in this work can be directly extended to any active learning problem that can be formalized using a hypothesis class, e.g., multiclass classification.}
We focus on the \textit{realizable} setting in this work, namely, there exists a target hypothesis $h^*\in \cH$ correctly labels every point.
In the pool-based active learning setting \cite{Hanneke14survey}, given a collected unlabeled data set $X = \{x_1,\ldots,x_m\}$, the learning goal is to recover the labels of $X$. 
Now just suppose the pool of unlabeled data $x_1,\ldots,x_m$ is available.
The possible labelings of these points form a subset of $\{0,1\}^m$, called the effective hypothesis class, which is 
\[\hat H = \{h(X)|h\in \cH\}\,,\]
where $h(X) = (h(x_1),\ldots, h(x_m))$ is the labeling of $X$ by $h$. Note that $|\hat H|\leq 2^m$
and $|\hat H|= \cO(m^d)$ if the VC dimension of $\cH$ is $d$.

In this work, we focus on the Bayesian setting~\cite{dasgupta2004analysis}, where
the target hypothesis is chosen in advance 
from some prior distribution $\pi$ over $\hat H$. Namely, without any additional information, for any labeling $h\in \hat H$, the probability that $h$ is the correct labeling of $X$ is $\pi(h)$. Since we can eliminate any hypothesis $h$ with $\pi(h)=0$ before starting to query for labels, we assume w.l.o.g. that $\pi(h)>0$ for all effective hypotheses in $\hat H$.

We remark that assuming that the unlabeled data $X$ is collected from some distribution $D_x$, which is essentially a distribution $D$  projected onto its input space, and that this distribution $D_x$ can be accurately classified by a hypothesis in $H$ with VC dimension $d$, standard generalization guarantees apply when the prior $\pi$ over $H$ is uniform (see \cite{dasgupta2004analysis} for more details).% I.e., for any $\epsilon>0$ if $m=\O(1/\epsilon)$, the maximal geralization error will be $\leq \epsilon$ w.h.p..}

\textbf{Standard active learning model} In the standard pool-based active learning setting, 
a single agent owns the pool of unlabeled data $X$. 
The agent, who knows both $\hat H$ and $\pi$, can query the labels of points in $X$, and her goal is to recover the labeling of $X$ (or to find the target hypothesis) by querying as few points as possible. 

A \textit{standard query algorithm} receives as input the prior distribution $\pi$ and unlabeled data set, $X$. In each iteration  $t=1,2,\ldots$, given the history up to time $t$, $$\cF_t =((x_1,y_1),\ldots,(x_{t-1},y_{t-1}))\in (X \times \{0,1\})^{t-1},$$ it selects a point $x_t$ to query and observes its label, $y_t$.  
The algorithm stops when all the labels of $X$ are recovered. Alternatively, the algorithm stops when for every two hypotheses $h_1,h_2\in \hat H$ consistent with $\cF_t$ (meaning that $h_1(x_\tau) = h_2(x_\tau) = y_\tau$ for all $\tau=1,\ldots, t-1$), $h_1(X)=h_2(X)$.

\textbf{Collaborative active learning model.}
In the collaborative setting, we assume there is more than one agent.  
Formally, there are $k$ agents and each agent $i$ has an individual unlabeled data set $X_i$ such that they together compose the pool, i.e., $\cup_{i\in [k]}X_i = X$, and each can query points from their own set $X_i$ (but cannot query points which are not in their set).
The goal of each agent is to recover the true labeling of their own set while performing as few queries as possible. 
The collaboration protocol, also called principal, who knows $\xset$, $\hat{H}$ and $\pi$, decides which point should be queried at each iteration, and her goal is to recover all the labels of $X$ using as few queries as possible. 
We remark that since data points belong to agents, queries of any point $x\in X$ can only be performed by agents whose data set contains $x$.

The query algorithm in the collaborative setting is similar to that in the standard setting, except that the algorithm needs to coordinate among the agents and decide which agent will query each point as some data points might belong to more than one agent.
In this setting, agents can decide to join the collaboration or learn individually at the beginning of the learning. But if they join the collaboration, they commit to follow the instructions of the query algorithm.
Therefore, given a prior distribution $\pi$ over $\hat H$ and a set of agents who would join the collaboration, w.l.o.g. denoted as $\{X_1,\ldots,X_\kappa\}$ for some $\kappa\in [k]$, at time $t=1,2,\ldots$, a \textit{collaborative query algorithm} asks agent $i_t\in [\kappa]$ such that $x_t\in X_{i_t}$ to query point $x_t$, and 
observes its label, $y_t$; the algorithm stops when the labels of points in $\cup_{i\in [\kappa]}X_i$ are completely recovered.

It is straightforward to check that standard query algorithms are a special case of collaborative query algorithms when there is a single agent, i.e., $k=1$. Additionally, a standard algorithm can also be run over multiple agents by considering the union of their data $\cup_{i\in [\kappa]} X_i$ as a single agent.
Hence, we omit ``standard'' or ``collaborative'' in a query algorithm when it is clear from the context how many agents are involved.

For any collaborative algorithm $\cA$, given an input $\pi$ and any collection of unlabeled data sets $X_1,\ldots,X_\kappa\subseteq X$ of size $\kappa\geq 1$, we denote by $Q(\cA,\pi,\{X_1,\ldots,X_\kappa\}, h)$ the label complexity (number of label queries) of $\cA(\pi,\{X_1,\ldots,X_\kappa\})$ when the target hypothesis is $h$. 
For randomized algorithms, the label complexity is taken expectation over the randomness of the algorithm. 
We define the label complexity as follows.
\begin{definition}[Label complexity] 
Given any fixed unlabeled pool and effective hypothesis class $(X,\hat H)$, for any algorithm $\cA$, prior distribution $\pi$ over $\hat H$ and any collection of unlabeled data sets $X_1,\ldots,X_\kappa\subseteq X$ of size $\kappa\geq 1$, 
the label complexity of $\cA$ with $(\pi, \{X_1,\ldots,X_\kappa\})$ as input, denoted by $Q(\cA,\pi, \{X_1,\ldots,X_\kappa\})$, is the expected number of label queries when $h$ is drawn from the prior $\pi$, i.e.,
    \[Q(\cA, \pi, \{X_1,\ldots,X_\kappa\}) = \EEs{h\sim \pi}{Q(\cA, h, \{X_1,\ldots,X_\kappa\})}\,.\]
    For each agent $i\in [\kappa]$ in the collaboration, we let $Q_i(\cA, \pi, \{X_1,\ldots,X_\kappa\})$ denote the expected number of queries performed by agent $i$.
\end{definition}
For any $(\pi, \{X_1,\ldots,X_\kappa\})$, let $Q^*(\pi, \{X_1,\ldots,X_\kappa\})=\min_{\cA} Q(\cA,\pi, \{X_1,\ldots,X_\kappa\})$ denote the optimal query complexity.
An algorithm $\cA$ is said to be \textit{optimal} if $Q(\cA, \pi,\{X_1,\ldots,X_\kappa\}) = Q^*(\pi, \{X_1,\ldots,X_\kappa\})$ for any prior distribution $\pi$ and $X_1,\ldots,X_\kappa$.

\textbf{Rational agents}
We assume that agents have access to a baseline algorithm and are able to run it on their own local data. 
Agents can decide to join the collaboration or run the baseline individually at the beginning of the learning.
If they join the collaboration, they commit to follow the instructions of the query algorithm. 
Each agent is incentivized to join the collaboration if she could perform fewer label queries (assuming that all others join the collaboration) by pulling out and running the baseline $\cA$ individually. 
Formally, 
\begin{definition}[Individual rationality]
    In a collaborative learning problem with prior distribution $\pi$ and $k$ agents $\xset$, given a baseline algorithm $\cA$,
    a collaborative algorithm $\cA'$ is individually rational (IR) if 
\begin{equation}
    {Q_i(\cA',\pi, \xset)} \leq Q(\cA, \pi, \{X_i\}), \forall i\in [k]\,.\label{eq:def-incentive}
\end{equation}
\end{definition}
We say 
 $\cA'$ is \textit{strictly individually rational} (henceforth, SIR) if
\begin{equation*}
    {Q_i(\cA',\pi, \xset)} < Q(\cA, \pi, \{X_i\}), \forall i\in [k]\,.
\end{equation*}
We remark that in addition to their own sets, each agent knows all the unlabeled data sets,  $\xset$, and the prior distribution $\pi$, otherwise, they will not be able to compute $Q_i(\cA',\pi,\xset)$. The principal also has access to $\xset$ and $\pi$, and can therefore make sure these constraints are satisfied.

It is worth noting that our model can also accommodate different individual baseline algorithms. We briefly discuss this in Section~\ref{sec:discussion}.

An alternative interpretation of the problem in a game theoretic framework is as follows: each agent has a strategy space of two strategies, joining the collaboration and not.
The utility of an agent that performs $Q$ queries is $-Q$.
If the algorithm $\cA$ is IR, then the case of all agents joining the collaboration is a Nash equilibrium (since switching to not joining will not increase their utility).
If $\cA$ is SIR, then all agents joining the collaboration is a strict Nash equilibrium.

\section{Construction of IR Collaborative Algorithms}\label{sec:ir}
When agents are limited to a poor baseline algorithm, e.g., randomly selecting points to query, the principal can simply incentivize agents to collaborate by using a superior algorithm that requires fewer labeling efforts. 
We therefore start by considering optimal baseline algorithms in Section~\ref{subsec:ir-opt}.
If we are able to find an IR collaborative algorithm for an optimal baseline algorithm, $\OPT$, then it must be IR w.r.t. all baseline algorithms.
We demonstrate that, surprisingly, the optimal algorithm $\OPT$ is IR given that the baseline algorithm is $\OPT$ itself.
Since computing an optimal algorithm is known to be NP-hard, we continue by considering the best-known approximation algorithm, the greedy algorithm. 
In Section~\ref{subsec:ir-greedy}, we show that given the greedy algorithm as baseline, the collaboration protocol that runs the greedy algorithm is not IR. 
Then in Section~\ref{subsec:ir-alg}, 
we provide a general scheme that transforms any baseline algorithm into an IR algorithm while maintaining a comparable label complexity.
\subsection{Optimality Implies Universal Individual Rationality}\label{subsec:ir-opt}
Incorporating individual rationality as an additional constraint to optimality usually requires additional effort in certain settings, e.g., in online learning by \cite{blum2020advancing}. 
However, in our specific setting, optimality does not contradict the individual rationality property. That is, an optimal algorithm will not increase any agent's label complexity to benefit other agents. 
In fact, optimizing for optimality implies achieving individual rationality for all baseline algorithms. 
\begin{theorem}\label{thm:opt-ir}
For any optimal collaborative algorithm $\OPT$, we have $${Q_i(\OPT,\pi, \xset)} \leq Q(\OPT, \pi, \{X_i\})= Q^*(\pi, \{X_i\}), \forall i\in [k].$$
    Therefore, $\OPT$ is IR w.r.t. any baseline algorithm.
\end{theorem}

We prove the theorem by contradiction. If $\OPT$ is not IR for the baseline being $\OPT$, then there exists an agent $i$ such that $Q(\OPT, \pi, \{X_i\}) < Q_i(\OPT,\pi, \xset)$. In this case, we can construct a new algorithm by first running $\OPT$ over $\{X_i\}$ (to recover the labels of $X_i$)  and then running $\OPT(\pi, \xset\})$ and replacing agent $i$'s queries with the recovered the labels of $X_i$. This new algorithm incurs a strictly smaller label complexity than $\OPT$, which is a contradiction to the optimality of $\OPT$. The formal proof is deferred to Appendix~\ref{app:opt-ir}. 
Unfortunately, computing an optimal query algorithm is not just NP-hard, but also hard to approximate within
a factor of $\Omega(\log(|\hat H|))$~\cite{golovin2010near,chakaravarthy2007decision}. 
One of the most popular heuristics to find an approximated solution is greedy.
\subsection{The Greedy Algorithm is Not Individually Rational}\label{subsec:ir-greedy}
\vspace{-0.5em}

For standard Bayesian active learning, \cite{kosaraju2002optimal,dasgupta2004analysis} presented a simple greedy algorithm called generalized binary search (GBS), which chooses a point leading to the most balanced partition of the set of hypotheses consistent with the history. 
More specifically, at time step $t$, given the history $\cF_t =((x_1, i_1,y_1),\ldots,(x_{t-1},i_{t-1},y_{t-1}))$, let $\VS(\cF_t) = \{h\in \hat H|h(x_\tau) = y_\tau,\forall \tau\in [t-1]\}$ denote the set of hypotheses consistent with the history $\cF_t$ (often called the version space associated with $\cF_t$).
Given $\cF_t$ and $(\pi, \{X_1,\ldots,X_\kappa\})$ as input, GBS will query 
\[x_t = \argmax_{x\in \cup_{i\in [\kappa]}X_i} \min(\pi(\{h\in \VS(\cF_t)|h(x)=1\}), \pi(\{h\in \VS(\cF_t)|h(x)=0\}))\,\]
at time $t$. When referring to GBS as a collaborative algorithm, we complement it with an arbitrary tie-breaking rule for selecting $i_t$, as GBS itself does not specify how to choose which agent to query.
GBS is guaranteed to achieve competitive label complexity with the optimal label complexity.

\begin{lemma}[Optimality of GBS, Theorem~3 of \cite{dasgupta2004analysis}]\label{lemma:gbs}
    For any prior distribution $\pi$ over $\hat H$ and $k$ agents $\xset$, the label complexity of GBS satisfies that
    \[Q(\gbs,\pi,\xset) \leq 4Q^*(\pi,\xset) \ln(\frac{1}{\min_{h\in \hat H} \pi(h)})\,.\]
\end{lemma}
The greedy algorithm GBS not only achieves approximately optimal label complexity, but it is also computationally efficient, with a running time of $\bigO(m^2|\hat H|)$. 
As GBS is the best-known efficient approximation algorithm, it is natural to think that agents would adopt GBS as a baseline. 

As we have shown that the optimal algorithm is IR w.r.t. itself, the next natural question is: \textit{Is GBS (as collaboration protocol) individually rational w.r.t. GBS itself? }

We answer this question negatively, even in the case of two agents.
Even worse, we present an example in which an agent's label complexity is $\Omega(n)$ when participating in the collaboration, but only $\bigO(1)$ when not participating.

\begin{restatable}{theorem}{greedyir}\label{thm:greedyir}
For the algorithm of GBS, there exists an instance of $(X_1,X_2, \pi)$ with $|X_1|=n$, in which agent $1$ incurs a label complexity of  $Q_1(\gbs,\pi, \{X_1,X_2\}) = \Omega(n)$ when participating the collaboration and can achieve $Q(\gbs,\pi, \{X_1\}) = \bigO(1)$ when not participating. %
\end{restatable}
Intuitively, at each time step, GBS only searches for an $x_t$ which leads to the most balanced partition of the version space, which does not necessarily lead to the optimal point to query. Given additional label information from the other agent, GBS possibly choose a worse point to query.
In addition, the label complexity of GBS is upper bounded by the optimal label complexity multiplied by a logarithmic factor.  It is possible that the agent achieves a smaller multiplicative factor by running GBS individually and a larger factor in the collaboration.
To prove the theorem, we construct an instance in which there exists a hypothesis with a prior probability of $1/4$, such that if GBS runs on $\{X_1,X_2\}$ and this hypothesis is the target, GBS will query almost all the points in $X_1$ (in a particular order) before returning this hypothesis. We show this part by induction. Additionally, we compute the query tree by running GBS solely on $X_1$ and use it to show that in this case, GBS has an expected query complexity of $O(1)$.
The full construction of the instance and the proof of Theorem~\ref{thm:greedyir} is deferred to Appendix~\ref{app:greedyir}.

\subsection{A Scheme of Converting Algorithms to IR Algorithms}\label{subsec:ir-alg}
\vspace{-0.5em}

Given that the greedy algorithm has been proven to be not individually rational w.r.t. itself, we raise the following question: \emph{Is it possible to develop a general scheme that can generate an IR algorithm given any baseline algorithm?} In this section, we propose such a scheme that addresses this question. 
Moreover, given a baseline algorithm $\cA$, the resulting IR algorithm can achieve a label complexity comparable to implementing the baseline algorithm over all agents, i.e., $\cA(\pi, \xset)$.
It is important to note that we aim for the label complexity to be comparable to $Q(\cA,\pi, \xset)$ rather than $\sum_{i\in [k]} Q(\cA,\pi,{X_i})$, as the latter holds true by individual rationality.
Given an efficient approximately optimal algorithm as baseline (e.g., GBS), our scheme can provide an algorithm that simultaneously exhibits individual rationality, efficiency, and approximately optimal label complexity. 

For any baseline algorithm $\cA$, we define a new algorithm $\ir(\cA)$, which runs $\cA$ as a subroutine.
Basically, we first calculate the label complexity of agent $i$ both when she is in collaboration with all the other agents, i.e., $Q_i(\cA,\pi,\xset)$,  and when she is not in collaboration, i.e., $Q(\cA,\pi,\{X_i\})$, for all $i\in [k]$. 
By doing so, we can distinguish which agents can benefit from collaboration when running $\cA$ and which cannot. 
We denote the set of agents who cannot benefit from collaboration with all others when running $\cA$ as $S = \{i|Q_i(\cA, \pi, \xset)>Q(\cA,\pi,\{X_i\})\}$. 
For those who do not benefit from the collaboration, we just run $\cA$ on their own data.
For those who benefit from collaborating  with the others together, we run $\cA$ over all agents $[k]$-- 
Only whenever $\cA(\pi,\xset)$ asks to query the label of a point belonging to some $i\in S$, since we already recovered the labels of $X_i$, we just  feed $\cA(\pi,\xset)$ with this label without actually asking agent $i$ to query.  
The detailed algorithm is described in Algorithm \ref{alg:basic2ir}.

\begin{algorithm}[H]\caption{$\ir$}\label{alg:basic2ir}
    \begin{algorithmic}[1]
    \STATE \textbf{input:} A query algorithm $\cA$, set $\xset$ and prior $\pi$ over $\hat H$
    \STATE For each $i\in[k]$, calculate $Q_i(\cA,\pi,\xset)$  and $Q(\cA, \pi, \{X_i\})$.
    \STATE Let $S \leftarrow \{i|Q_i(\cA, \pi, \xset)>Q(\cA,\pi,\{X_i\})\}$ and $X_S\leftarrow \cup_{i\in S}X_i$ // the agents who do not benefit from collaboration
    \STATE \textbf{for} {each $i\in S$} \textbf{do} $Y_i\leftarrow$ Run $\cA$ over $\{X_i\}$ // recover the labels for agent $i$
    % \ENDFOR
    \FOR{$t=1,\ldots$}
        \STATE $(i_t,x_t)\leftarrow$ the querying agent and the query point from $\cA(\pi, \xset)$
        \STATE \textbf{if} {$i_t\in S$} \textbf{then} Feed the label of $x_t$ from $Y_{i_t}$// we already recovered the labels of $ X_{S}$
        \STATE \textbf{else}
        % \ELSE 
         Ask  agent $i_t$
        % $i\notin S$ who has $x_t\in X_i$ 
        to  query the label of $x_t$         
        % \ENDIF
    \ENDFOR
    \end{algorithmic}
\end{algorithm}

\begin{theorem}\label{thm:ir}
    For any baseline algorithm $\cA$, the algorithm $\ir(\cA)$ satisfies the following properties:
    \begin{itemize}[topsep=0pt,itemsep =-0.5ex,  leftmargin = 6mm]
        \item \textbf{IR property:} $\ir(\cA)$ is individually rational w.r.t. the baseline algorithm $\cA$.
        \item \textbf{Efficiency:} $\ir(\cA)$ runs in 
        $\bigO(k T_{\cA, Q} + m T_{\cA,0})$ time, where $T_{\cA,0}$ is the time of computing $(i_t,x_t)$ at each time $t$ for $\cA$ and $T_{\cA, Q}$ is the maximum time of computing $Q_i(\cA,\pi,\{X_1,\ldots,X_\kappa\})$ for an  agent $i$, unlabeled data $\{X_1,\ldots,X_\kappa\}$, and algorithm $\cA$.
        \item \textbf{Label complexity:} $Q(\ir(\cA), \pi,\xset)\leq Q(\cA, \pi, \xset)$.
    \end{itemize} 
\end{theorem}

The proof follows the algorithm description immediately.
Note that when the baseline is GBS, we have $T_{\gbs,0} = \cO(m)$. We can compute $Q_i(\cA,\pi,\{X_1,\ldots,X_\kappa\})$ by simulating over all effective hypotheses $h\in \hat H$. For each $h$, we will query at most $m$ rounds.
Therefore, we have $T_{\gbs,Q} = \cO(m^2|\hat H|)$ and we can run $\ir(\gbs)$ in $\bigO(k m^2 |\hat H|)$ time. 
Using GBS as the baseline, we derive the following corollary.
    
\begin{corollary}\label{crl:ir-greedy}
Given GBS as the baseline, $\ir(\gbs)$ is IR; runs in $\bigO(k m^2 |\hat H|)$ time; and satisfies that $Q(\ir(\gbs), \pi, \xset)\leq 4Q^*(\pi,\xset) \ln(\frac{1}{\min_{h\in \hat H} \pi(h)})$.
\end{corollary}

\section{Converting  Algorithms to SIR Algorithms}\label{sec:sir}
In Section~\ref{sec:ir},
we provided a generic scheme for constructing an IR algorithm given any baseline algorithm.
In this section, we focus on constructing SIR algorithms given IR algorithms.
Since strict individual rationality requires that agents strictly benefit from collaboration, this is impossible without further assumptions. 
For example, consider a set of agents who only have one single independent point in their own sets 
and a prior distribution that is uniform over all labelings. 
In this case, each agent, regardless of whether she collaborates or not, has a label complexity of $1$ and cannot \emph{strictly} benefit from collaboration as the other agents cannot obtain information about her data.

Now, let us consider a notion weaker than SIR, called $i$-partially SIR, in which only agent $i$ strictly benefits from the collaboration, and any other agent $j\neq i$ does not get worse by joining the collaboration.
More formally, 
\begin{definition}[Partially SIR algorithms]
    For any baseline algorithm $\cA$, for all $i\in [k]$, an algorithm $\cO_i$ is $i$-partially SIR, if $\cO_i$ satisfies that
    \[Q_i(\cO_i, \pi, \xset) < Q(\cA,\pi, \{X_i\})\,,\]
    and 
    \[Q_{j}(\cO_i,\pi,\xset)\leq Q(\cA,\pi,\{X_j\}), \forall j\in [k]\setminus \{i\}\,.\]
\end{definition}
If we are given an $i$-partially SIR algorithm $\cO_i$ for each $i$, then we can construct a SIR algorithm by running a mixture of an IR algorithm $\cA'$ (e.g., $\ir(\cA)$ in Algorithm~\ref{alg:basic2ir}) and $\{\cO_i|i\in [k]\}$ with the label complexity a little (arbitrarily small) higher than that of $\cA'$.
\begin{lemma}\label{lmm:partsir}
    For any baseline algorithm $\cA$, given an IR algorithm $\cA'$ and partially SIR algorithms $\{\cO_i|i\in [k]\}$, for any $\epsilon>0$, let $\cA''_\epsilon$ be the algorithm of running $\cA'$ with probability $(1-\frac{\epsilon}{n})$ and running $\cO_i$ with probability $\frac{\epsilon}{kn}$.
    Then $\cA''_\epsilon$ satisfies the following properties.
    \begin{itemize}[topsep=0pt,itemsep =-0.5ex,  leftmargin = 6mm]
        \item SIR property: $\cA''_\epsilon$ is SIR with respect to the baseline algorithm $\cA$.
        \item Label complexity: $Q(\cA''_\epsilon, \pi, \xset)\leq Q(\cA', \pi, \xset) +\epsilon$.
    \end{itemize}
\end{lemma}
The proof is straightforward from the definition, and we include it in Appendix~\ref{app:part-sir} for completeness.
Since a SIR algorithm is also $i$-partially SIR for all $i\in [k]$, constructing a SIR algorithm is equivalent to constructing a set of partially SIR algorithms $\{\cO_i|i\in [k]\}$.
Therefore, the problem of constructing a SIR algorithm is reduced to constructing partially SIR algorithms $\{\cO_i|i\in [k]\}$. 

For the remainder of this section, we will present the SIR results for an optimal baseline algorithm in Section~\ref{subsec:sir-opt}, where we propose a sufficient and necessary assumption for the existence of SIR algorithms and then provide a SIR algorithm.
This algorithm is SIR w.r.t. any baseline algorithm but again, computationally inefficient.
In Section~\ref{subsec:sir-apprx}, we provide a general scheme that transforms any baseline algorithm into a SIR algorithm.

\subsection{A Universal SIR Algorithm for Any Baseline Algorithm}\label{subsec:sir-opt}
\vspace{-0.5em}
Constructing a universal SIR algorithm w.r.t. any baseline is equivalent to constructing a SIR algorithm for an optimal baseline.
For the existence of SIR algorithms given any optimal baseline algorithm, we propose the following assumption, which is sufficient and necessary. We include the proof for the necessity of this assumption in Appendix~\ref{app:nece-asp}. The sufficiency of this assumption will be verified immediately after we construct a SIR algorithm.
\begin{assumption}\label{asp:clb-help}
We assume that for any $i\in [k]$, the optimal label complexity of agent $i$ given the information regarding the labels of all other agents is strictly smaller than that without this additional information, i.e.,
    $Q^*(\pi, \{X_i\})- \EEs{h\sim \pi}{Q^*(\pi_{h,-i}, \{X_i\})} >0\,,$
    where $\pi_{h,-i}$ is the posterior distribution of $\pi$ after observing $\{(x,h(x))|x\in \cup_{j\neq i} X_{j}\}$. 
\end{assumption}

According to Lemma~\ref{lmm:partsir}, we can construct a SIR algorithm by constructing a set of partially SIR algorithms $\{\cO_i|i\in [k]\}$. 

Let $\cO_i$ be the algorithm of running an optimal algorithm $\OPT$ over $(\pi, \{X_j|j\neq i\})$ first, then given the query-label history of $\{(x, h(x))|x\in \cup_{j\neq i} X_j\}$ for some $h\in \hat H$, run $\OPT$ over $(\pi_{h,-i}, \{X_i\})$. Then it immediately follows that $\cO_i$ is $i$-partially SIR from Assumption~\ref{asp:clb-help}. Let $\OPT_\epsilon''$ denote the algorithm of of running $\OPT$ with probability $(1-\frac{\epsilon}{n})$ and running $\cO_i$ with probability $\frac{\epsilon}{kn}$ for all $i\in [k]$. By Lemma~\ref{lmm:partsir}, we have

\begin{corollary}
    Under Assumption~\ref{asp:clb-help}, for any $\epsilon>0$, $\OPT_\epsilon''$ is SIR w.r.t. $\OPT$ and satisfies $$Q(\OPT''_\epsilon, \pi, \xset)\leq Q^*(\pi, \xset) +\epsilon.$$
    In addition, $\OPT_\epsilon''$ is SIR w.r.t. any baseline algorithm $\cA$ as $$Q_i(\OPT''_\epsilon, \pi, \xset)< Q^*(\pi, \{X_i\}) \leq Q(\cA,\pi,\{X_i\}).$$
\end{corollary}

\subsection{A Scheme of Converting Algorithms to SIR Algorithms
}\label{subsec:sir-apprx}
\vspace{-0.5em}
As mentioned before, computing an optimal algorithm is NP-hard. Assumption~\ref{asp:clb-help} assumes that collaboration can strictly benefit agents when the collaboration protocol can compute the optimal algorithm given $\pi_{h,-i}$.
Hence, the assumption does not take the computational issue into consideration and thus might not be enough for the existence of an efficient SIR algorithm w.r.t. an efficient approximation algorithm like GBS.

Instead, we propose prior-independent assumption that is sufficient for the existence of efficient SIR algorithms when we are given an efficient baseline and an efficient IR algorithm w.r.t. the baseline. 
Basically, we assume that, there exists an effective hypothesis $h\in \hat H$, given the information that all other agents are labeled by $h$, the number of labelings of $X_i$ consistent with the label information is strictly smaller than the total number of labelings of $X_i$ by $\hat H$. Formally, for any $i\in [k]$, let $X_{-i} = \cup_{j\neq i} X_j$ denote the union of all agents' data except agent $i$.
Let $H(X_i)=\{h'(X_i)|h'\in \hat H\}$ denote the effective hypothesis class of $X_i$, i.e., all labelings of $X_i$.
For any $h\in \hat H$, let $H(X|h) = \{h'(X_i)|h'(X_{-i}) = h(X_{-i}),h'\in \hat H\}$ denote the subset which are consistent with all other agents being labeled by $h$.

\begin{assumption}\label{asp:clb-help-efficient}
    For all $i\in [k]$, there exists an $h\in \hat H$ s.t.  
    the number of labelings of $X_i$ consistent with $(X_{-i}, h(X_{-i}))$ is strictly smaller than the number of labelings by $\hat H$, i.e., $\abs{H(X_i|h)} < \abs{H(X_i)}$.
\end{assumption}
Intuitively, Assumption~\ref{asp:clb-help-efficient} means that for every agent $i$, there exists an hypothesis  $h$ such that when $h^*=h$, the cardinality of the set of hypotheses consistent with $(X_{-i}, h(X_{-i}))$ is strictly smaller than $|\hat H|$. We will show that this assumption is sufficient for the \textit{existence} of algorithms satisfying SIR property. Without it, it is unclear if there exist SIR algorithms. 
The assumption can be easily verified by iterating each $h\in \hat H$ (this is polynomial in  $|\hat H|$ and $m$). 

Notice that each deterministic query algorithm $\cA$ can be represented as a binary tree, $\cT_\cA$ whose internal nodes at level $t$ are queries (``what is the $x_t$'s label?''), and whose leaves are labelings as illustrated in Figure~\ref{fig:query-tree}.
Under Assumption~\ref{asp:clb-help-efficient}, we can prune the query tree of $\cA(\pi, \{X_i\})$ by removing
all subtrees whose leaves are all in $H(X_i)\setminus H(X_i|h)$.
We do not need to construct this pruned tree when we implement the algorithm. At time $t$, we just need to generate an $x_t$ from $\cA(\pi, \{X_i\})$, then check if this node should be pruned by checking if all the hypotheses $H(X_i|h)$ agree on the label of $x_t$. If this is true, it means that we have already recovered the label of $x_t$ and thus we just need to feed the label to the algorithm without actually querying $x_t$ again.
Then we can construct a $i$-partially SIR algorithm $\cO_i$ by running $\ir(\cA)$ over $(\pi, \{X_j|j\neq i\})$ to recover the labeling of $X_{-i}$ first, then running pruned version of $\cA(\pi, \{X_i\})$. Note that the implementation also works when $\cA$ is randomized.

Consider Example~\ref{ex:treshold}, where the hypothesis class $\mathcal H = \{x\geq \alpha|\alpha=0.2,0.4,0.5,0.6,0.8\}$ with a uniform prior, agent $1$ with points $X_1=\{0.25,0.5,0.75\}$ and agent $2$ with points $X_2=\{0.3,0.45, 0.55, 0.7\}$. 
When agent $1$ runs a binary search, the query tree has $0.5$ as a root, then if $0.25$ if $h^*(0.5)=1$, and $0.75$  otherwise. Now, algorithm $\cO_1$ runs a binary search on $X_2$ and obtains all the labels of the points in $X_2$. The hypothesis $h = 1(x\geq 0.5)$ holds  $|H(X_1|h)| = 1 < |\hat H(X_1)|=4$. 
When $h$ is the labeling function, $0.3$ and $0.45$ are labeled as negative, and $0.55$ and $0.7$ are labeled as positive. Then, $\cO_1$ will only need agent $1$ to query $0.5$ as the labels of $0.25$ and $0.75$ can be inferred and they are pruned in the query tree. 

\begin{lemma}\label{lemma:OiIsI-Part-SIR}
Under Assumption~\ref{asp:clb-help-efficient}, the algorithm $\cO_i$ constructed above is $i$-partially SIR and runs in time $\bigO(\cT_{\ir(\cA)} + m (|\hat H| + T_{\cA,0}))$ time, where $\cT_{\ir(\cA)}$ is the running time of $\ir(\cA)$ and $T_{\cA,0}$ is the time of computing $(i_t,x_t)$ at each time $t$ for $\cA$.

\end{lemma}
The proof of Lemma~\ref{lemma:OiIsI-Part-SIR} is deferred to Appendix~\ref{app_lmm3}.

We can then construct an algorithm $\cA_\epsilon''$ by running $\ir(\cA)$ with probability with probability $(1-\frac{\epsilon}{n})$ and running $\cO_i$ constructed in the above way with probability $\frac{\epsilon}{kn}$ for all $i\in [k]$. By combining Lemmas~\ref{lmm:partsir} and \ref{lemma:OiIsI-Part-SIR}, we derive the following theorem. Then, combining it with Corollary~\ref{crl:ir-greedy}, we derive a SIR algorithm for GBS as baseline GBS.
\begin{theorem}
Under Assumption~\ref{asp:clb-help-efficient}, for any baseline algorithm $\cA$, for any $\epsilon>0$, $\cA_\epsilon''$ is SIR and satisfies $$Q(\cA''_\epsilon, \pi, \xset)\leq Q(\ir(\cA),\pi, \xset) +\epsilon.$$
In addition, Algorithm $\cA_\epsilon''$ runs in $\bigO(\cT_{\ir(\cA)} + m (|\hat H| + T_{\cA,0}))$ time.
\end{theorem}

\begin{corollary}
Given GBS as the baseline, Algorithm $\gbs_\epsilon''$ is SIR; runs in $\bigO(k m^2 |\hat H|)$ time; and satisfies that $$Q(\gbs_\epsilon'', \pi, \xset)\leq 4Q^*(\pi, \xset) \ln\left(\frac{1}{\min_{h\in \cH}\pi(h)}\right) +\epsilon.$$
\end{corollary}

\section{Discussion}\label{sec:discussion}
In this paper, we have initiated the study of collaboration in active learning in the presence of incentivized agents. We first show that an optimal collaborative algorithm is IR w.r.t. any baseline algorithm while approximate algorithms are not.
Then we provide meta-algorithms capable of producing IR/SIR algorithms given any baseline algorithm as input.

Our model and algorithms can also allow different agents to have different baseline algorithms---Whenever the principal plans to run an algorithm on a union of datasets, she can simply check which baseline algorithm has the lowest expected query complexity on this union, and run it. When she needs to run an algorithm on a dataset of an individual, she can simply run their baseline. This way the (S)IR is preserved.
 
There are a few problems we leave open. First, relaxing the assumption that each agent $i$ has full knowledge of $X_{-i}$ (e.g., due to privacy concerns). Second, relaxing the assumption that agents provide reliable labels. 
% First, designing IR algorithms for stream-based active learning, or  relaxing the assumption that each agent has full knowledge of $X_{-i}$ (e.g., due to privacy concerns). 
Third, deriving results in non-realizable settings. %allowing different agents to have different baseline algorithms and be competitive with all of them (e.g., perhaps some agents care about different performance metrics such as average or worst case running time)
Realizability is a standard assumption in learning theory at large and particularly within active learning as highlighted in the classical machine learning theory textbook by~\cite{Shalev-Shwartz_Ben-David_2014} and the active learning theory survey by~\cite{Hanneke14survey}. 
Moreover, realizability has also been adopted in the collaborative learning setting (e.g.,~\cite{Blum17}). The reason is that without realizability, additional complications might arise in collaboration such as why collaboration would yield benefits. In general, we believe that additional assumptions would be required to relax this assumption. Forth, towards a more game theory orientation, it would be interesting to design collaborative algorithms in a setting where agents can form coalitions. 
Finally, finding a necessary and sufficient assumption(s) for the existence of efficient SIR algorithms will be an interesting direction (and we have found a sufficient one in this work).

\section*{Acknowledgements}
We would like to thank Avrim Blum for several useful discussions.
This work was supported in part by the National Science Foundation under grants 2212968 and 2216899, by the Simons Foundation under the Simons Collaboration on the Theory of Algorithmic Fairness, by the Defense Advanced Research Projects Agency under cooperative agreement HR00112020003. The views expressed in this work do not necessarily reflect the position or the policy of the Government and no official endorsement should be inferred. Approved for public release; distribution is unlimited.

\bibliographystyle{apalike}
\bibliography{ref}

\newpage
\appendix
\section{Proof of Theorem~\ref{thm:opt-ir}}\label{app:opt-ir}
\begin{proof}
    For any randomized optimal algorithm $\opt$, every realization of the internal randomness of $\opt$ must have the same query complexity and be optimal.
    Otherwise, there exists a realization with query complexity smaller than $\opt$, which conflicts with that $\opt$ is optimal. 
    Therefore, it suffices to prove the theorem for deterministic optimal collaborative algorithms.
    
    We will prove the theorem for deterministic algorithms by contradiction.
    Suppose that there exists a deterministic optimal collaborative algorithm $\opt$ that is not individually rational w.r.t. running itself as baseline. Hence, there exists an agent $i\in [k]$ such that $Q_{i}(\opt,\pi, \xset) > Q(\opt, \pi, \{X_i\})$.   
    In this case, we can construct another algorithm $\cA'$ with smaller label complexity, which will contradict  the optimality of $\opt$.
    
    The basic idea of $\cA'$ is to run $\opt$ over $(\pi,\{X_i\})$ first and to recover the labels of $X_i$. 
    Then, $\cA'$ simulates $\opt(\pi, \xset)$ and asks $\opt(\pi, \xset)$ what point to query. But whenever $\opt(\pi, \xset)$ asks to query the label of some point in $X_i$, since we already know the labeling of $X_i$, we can just feed $\opt(\pi, \xset)$ with these labels without actually asking agent $i$ to query them. 
    
    Thus, the label complexity of $\cA'$ is
    \begin{align*}
        Q(\cA',\pi,\xset) &= Q(\opt, \pi, \{X_i\}) + \sum_{j:j\neq i}Q_j(\opt, \pi, \xset)\\
        &< Q_{i}(\opt,\pi, \xset) + \sum_{j:j\neq i}Q_j(\opt, \pi, \xset) \\
        &= Q(\opt, \pi, \xset)= Q^*(\pi, \xset)\,,
    \end{align*}
    where the first inequality holds due to that $\opt$ is not IR and the last equality holds since $\opt$ is optimal.
    Since $Q^*(\pi, \xset)\leq Q(\cA',\pi,\xset)$ by definition, there is a contradiction.
\end{proof}
\section{Proof of Theorem~\ref{thm:greedyir}}\label{app:greedyir}
% \greedyir*
\begin{proof}
    The construction is inspired by~\cite{dasgupta2004analysis}.
    Consider $k=2$ and let the unlabeled data set of agent $1$ be 
    \[X_1 = \{(0,1,0),(0,2,0), (0,0,1), (0,0,2),\ldots, (0,0,n)\}\]
    for some $n\in \NN_+$.
    
    Let  the unlabeled data set of 
    agent $2$ be
    \[X_2 = \{(1,0,0)\}\,.\]
    Let the unlabeled pool $X = X_1\cup X_2$.
    Let $h_{i,j,l}$ denote the hypothesis which labels $(i,0,0),(0,j,0),(0,0,l)$ as $1$ and the rest as $0$.
    
    Let the hypothesis class be $\cH = \{h_{i,j,l}|i\in \{0,1\}, j\in [2], l\in [n]\}$.
    
    Let the prior distribution $\pi_0 = \pi$ be defined as follows:
    \begin{equation*}
        \begin{cases}
            \pi(h_{0,0,0}) = \frac{1}{4}&\\
            \pi(h_{0,j,l}) = \frac{1}{4 \cdot 3^l}& \text{ for } j=1,2, l=1,\ldots,n-1\\
            \pi(h_{0,j,n}) = \frac{1}{8\cdot 3^{n-1}}& \text{ for } j=1,2\\
            \pi(h_{1,1,l}) = \frac{1}{3^l}& \text{ for } l=1,\ldots,n-1\\
            \pi(h_{1,1,n}) = \frac{1}{2\cdot 3^{n-1}}\,.&
        \end{cases}
    \end{equation*}

    Now we show that the label complexity of agent $1$ in the collaboration is $Q_1(\gbs, \pi, \{X_1,X_2\}) = \Omega(n)$. While the label complexity of running GBS itself is $Q(\gbs,\pi, \{X_1\}) = \bigO(1)$.

    \paragraph{Label complexity of agent $1$ in the collaboration} Let $\VS$ denote the version space. And for any point $x$, let $\VS_{x}^+ = \{h\in \VS|h(x) =1\}$ denote the subset of the version space which labels $x$ by $1$. Similarly, let $\VS_{x}^- = \{h\in \VS|h(x) =0\}$.

    Now let us consider the length of the %query
    path in the query tree when the target hypothesis is $h_{0,0,0}$.
    
    A-priori (before starting to query), for point $(1,0,0)$, we have 
    \begin{equation*}
        \pi(\VS_{(1,0,0)}^+) = \sum_{l=1}^n \pi(h_{1,1,l}) = \frac{1}{3} +\frac{1}{3^2} +\ldots +\frac{1}{3^{n-1}} +\frac{1}{2\cdot 3^{n-1}} =\frac{1}{2}\,.
    \end{equation*}   
    For point $(0,1,0)$, we have
    \[\pi(\VS_{(0,1,0)}^+) = \sum_{l=1}^n (\pi(h_{0,1,l}) + \pi(h_{1,1,l}))= \sum_{l=1}^n \pi(h_{0,1,l}) + \pi(\VS_{(1,0,0)}^+) >\frac{1}{2}\,.\]
    For point $(0,2,0)$, we have
    \[\pi(\VS_{(0,2,0)}^+) = \sum_{l=1}^n \pi(h_{0,2,l}) = \frac{1}{4\cdot 3} +\frac{1}{4\cdot 3^2} +\ldots +\frac{1}{4\cdot 3^{n-1}} +\frac{1}{8\cdot 3^{n-1}} =\frac{1}{8}\,.\]
    For other points $(0,0,l)$ for $l\in [n-1]$, we have $\pi(\VS_{(0,0,l)}^+) = \frac{1}{4\cdot 3^l} + \frac{1}{3^l} = \frac{5}{4\cdot 3^l}<\frac{1}{2}$ and for $(0,0,n)$, we have $\pi(\VS_{(0,0,n)}^+)<\pi(\VS_{(0,0,n-1)}^+)<\frac{1}{2}$.
    
    Therefore, the algorithm $\gbs(\pi,\{X_1,X_2\})$ will query $(1,0,0)$ at time $1$.
    Suppose the label of $(1,0,0)$ is $0$ since we consider the path corresponding to $h_{0,0,0}$ as the target hypothesis.
    
    Now we show that $\gbs(\pi,\{X_1,X_2\})$ will query points $(0,0,1), (0,0,2),\ldots, (0,0,n)$ sequentially by induction.
    
    At time $1$, the version space is $\VS = \{h_{0,0,0}\}\cup \{h_{i,j,l}\in \hat H|i=0\}$. We list $\pi(h_{0,j,l})$ for $j\in \{1,2\}$ and $l\in [n]$ in Table~\ref{tab:pi-0jl} for illustration.
    \begin{table}[H]
        \centering
        \begin{tabular}{c|c|c}
             & $(0,1,0)$ & $(0,2,0)$\\\hline
            $(0,0,1)$ & $\frac{1}{4 \cdot 3}$ & $\frac{1}{4 \cdot 3}$\\\hline
            $(0,0,2)$ &$\frac{1}{4 \cdot 3^2}$ & $\frac{1}{4 \cdot 3^2}$ \\\hline
            $\cdots$ & $\cdots$ & $\cdots$ \\\hline
            $(0,0,n)$ & $\frac{1}{8 \cdot 3^{n-1}}$ & $\frac{1}{8 \cdot 3^{n-1}}$\\\hline
        \end{tabular}
        \caption{Table of $\pi(h_{0,j,l})$ for $j\in \{1,2\}$ and $l\in [n]$.}
        \label{tab:pi-0jl}
    \end{table}
    Then we can compute that 
    \[\pi(S_{(0,1,0)}^+) = \pi(S_{(0,2,0)}^+)=\frac{1}{4\cdot 3} +\frac{1}{4\cdot 3^2} +\ldots +\frac{1}{4\cdot 3^{n-1}} +\frac{1}{8\cdot 3^{n-1}} =\frac{1}{8}\,,\]
    \[\pi(S_{(0,0,1)}^+) = \frac{1}{6}> \pi(S_{(0,0,l)}^+)\,,\]
    \[\pi(S_{(0,0,l)}^+) \leq \pi(S_{(0,0,2)}^+) = \frac{1}{18}\,,\]
    for all $l\geq 2$. 
    
    Thus, the algorithm $\gbs(\pi,\{X_1,X_2\})$ will choose $(0,0,1)$ at time $2$. 
    
    Suppose that at time $t=2,3,\ldots,l$, $\gbs(\pi,\{X_1,X_2\})$ has picked $(0,0,1),\ldots, (0,0,l-1)$ and all are labeled $0$.
    
    Now we show that $\gbs(\pi,\{X_1,X_2\})$ will pick $(0,0,l)$ at time $t= l+1$.
    The version space at the beginning of time $l+1$ is
    $\VS = \{h_{0,0,0}\}\cup \{h_{i,j,p}\in \hat H|p\geq l\}$.
    We can compute that 
    \[\pi(S_{(0,0,l)}^+) = \frac{1}{2\cdot 3^l}> \pi(S_{(0,0,p)}^+)\] 
    for all $p>l$, and that 
    \[
    \pi(S_{(0,1,0)}^+) = \pi(S_{(0,2,0)}^+)= \frac{1}{4 \cdot 3^l} + \frac{1}{4 \cdot 3^{l+1}} + \ldots+\frac{1}{4 \cdot 3^{n-1}} + \frac{1}{8 \cdot 3^{n-1}} = \frac{1}{8 \cdot 3^{l-1}}\,.
    \]
    Hence, $\gbs(\pi,\{X_1,X_2\})$ will pick $(0,0,l)$.
    
    Therefore, we proved that when the target hypothesis is $h_{0,0,0}$, $\gbs(\pi,\{X_1,X_2\})$ will query \\
    $(1,0,0),(0,0,1), (0,0,2),\ldots, (0,0,n)$ sequentially. 
    
    Thus, we have that $Q_1(\gbs,\pi,\{X_1,X_2\},h_{0,0,0}) = n + 1$, and 
    $Q_1(\gbs,\pi,\{X_1,X_2\}) \geq \frac{n+1}{4}$ as $\pi(h_{0,0,0}) =\frac{1}{4}$.

\paragraph{Label complexity of agent $1$ when she runs the (GBS) baseline individually}
Now we show that $Q(\gbs,\pi,\{X_1\}) = \bigO(1)$. Since $X_1$ does not contain $(1,0,0)$, both $h_{0,j,l}$ and $h_{1,j,l}$ label $X_1$ identically. Every effective hypothesis over $X_1$ can be written as $h_{*,j,l}$ with $\pi(h_{*,j,l}) = \pi(h_{0,j,l})+\pi(h_{1,j,l})$, which is listed in Table~\ref{tab:pi-xjl}.
       \begin{table}[H]
        \centering
        \begin{tabular}{c|c|c}
             & $(0,1,0)$ & $(0,2,0)$\\\hline
            $(0,0,1)$ & $\frac{1}{4 \cdot 3} + \frac{1}{3}$ & $\frac{1}{4 \cdot 3}$\\\hline
            $(0,0,2)$ &$\frac{1}{4 \cdot 3^2}+ \frac{1}{3^2}$ & $\frac{1}{4 \cdot 3^2}$ \\\hline
            $\cdots$ & $\cdots$ & $\cdots$ \\\hline
            $(0,0,n)$ & $\frac{1}{8 \cdot 3^{n-1}} +\frac{1}{2 \cdot 3^{n-1}}$ & $\frac{1}{8 \cdot 3^{n-1}}$\\\hline
        \end{tabular}
        \caption{Table of $\pi(h_{*,j,l})$ for $j\in \{1,2\}$ and $l\in [n]$.}
        \label{tab:pi-xjl}
    \end{table}  
    Notice that if we know that the label of $(0,0,l)$ is positive for some $l$, then the version space has at most $2$ effective hypotheses, $h_{*,1,l}$ and $h_{*,2,l}$.
    In this case, the algorithm needs at most $2$ more queries.
    
    At time $t=1$, we have
    \[\pi(S_{(0,0,1)}^+) = \frac{1}{4 \cdot 3} + \frac{1}{3} + \frac{1}{4 \cdot 3} =\frac{1}{2}\,,\] 
    \[\pi(S_{(0,0,l)}^+) < \pi(S_{(0,0,1)}^+)\,,\forall l\geq 2\,,\]
    \[\pi(S_{(0,2,0)}^+) = \frac{1}{4\cdot 3} +\frac{1}{4\cdot 3^2} +\ldots +\frac{1}{4\cdot 3^{n-1}} +\frac{1}{8\cdot 3^{n-1}} = \frac{1}{8}\,,\]
    \[\pi(S_{(0,1,0)}^+) = \pi(S_{(0,2,0)}^+) \cdot 5 =\frac{5}{8}\,.\]
    Therefore, $\gbs(\pi,\{X_1\})$ will query $(0,0,1)$ at $t=1$.
    
    We complete the proof by exhaustion. 
    If $(0,0,1)$ is labeled as $1$, then the algorithm needs at most two more queries as aforementioned.
    
    If $(0,0,1)$ is labeled as $0$, then $h_{*,1,1}$ and $h_{*,2,1}$ will be removed from the version space and $\gbs(\pi,\{X_1\})$ will query $(0,1,0)$ at $t=2$ then.
    
    If the label is $1$, the version space is reduced to $\{h_{*,1,l}|l=2,\ldots,n\}$ and $\gbs(\pi,\{X_1\})$ will query $(0,0,2),(0,0,3),\ldots$ sequentially until receiving a positive label.
    
    If the label of $(0,1,0)$ is $0$, $\gbs(\pi,\{X_1\})$ will query $(0,2,0)$ at time $t=3$.
    If the label of $(0,2,0)$ is $1$, then it is similar to the case of  $(0,1,0)$ being labeled $1$ and the algorithm will query $(0,0,2),(0,0,3),\ldots$ sequentially.
    
    If the label of $(0,2,0)$ is $0$, we know the target hypothesis is $h_{0,0,0}$ and we are done.
    
    Hence we have $Q(\gbs,\pi,\{X_1\}) \leq \sum_{l=1}^n (\pi(h_{*,1,l}) + \pi(h_{*,2,l})) \cdot (3+ l) + \pi(h_{0,0,0})\cdot 3 = \sum_{l=1}^{n-1} \frac{1}{2\cdot 3^{l-1}}  \cdot (3+ l) + \frac{1}{4}\cdot 3 = \bigO(1)$.
\end{proof}
\section{Proof of Lemma~\ref{lmm:partsir}}\label{app:part-sir}
\begin{proof}
\textbf{SIR property: } Since $\cA'$ and $\{\cO_i|i\in [k]\}$ are IR and agent $i$ can strictly benefit from $\cO_i$, 
we have $Q_i(\cA_\epsilon'', \pi, \xset) < Q(\cA,\pi, \{X_i\})$ for all $i\in [k]$.

\textbf{Label complexity: } The label complexity of $\cA_\epsilon''$ is
\begin{align*}
    Q(\cA_\epsilon'',\pi, \xset) = & (1-\frac{\epsilon}{n})Q(\cA',\pi,\xset) + \frac{\epsilon}{kn} \sum_{i=1}^k Q(\cO_i,\pi,\xset)\\
    \leq& (1-\frac{\epsilon}{n})Q(\cA',\pi,\xset) + \epsilon\,.
\end{align*}
Then we are done.
\end{proof}
\section{Proof of necessity of Assumption~\ref{asp:clb-help}}\label{app:nece-asp}
\begin{proof}[Proof of necessity]
Suppose that there exists an SIR algorithm $\cA''$ when the baseline algorithm is optimal. We therefore have 
$Q_i(\cA'',\pi,\xset) < Q^*(\pi,\{X_i\})$ by definition.  
We claim that $\cA''$ must satisfy $Q_i(\cA'',\pi,\xset) \geq \EEs{h\sim \pi}{Q^*(\pi_{h,-i}, \{X_i\})}$.
This is because we can construct another algorithm $\cB$ by running $\cA''$ over all other agents except agent $i$, i.e., running $\cA''$ over $(\pi, X_{-i})$ with $X_{-i}= \{X_j|j\neq i\}$ first to recover the labels of all other agents $X_{-i}$.
Then, $\cB$ simulates $\cA''$ over $(\pi, \xset)$ without actually querying any point in $X_{-i}$ (similarly to Algorithm~\ref{alg:basic2ir}).
In this case, the label complexities of agent $i$ are identical for algorithms $\cB$ and $\cA''$, i.e., $Q_i(\cB,\pi,\xset)= Q_i(\cA'',\pi,\xset)$. 
Since $Q^*(\pi_{h,-i}, \{X_i\})$ is the optimal label complexity of agent $i$ given the label information of $X_{-i}$, we have $Q_i(\cB,\pi,\xset) \geq Q^*(\pi_{h,-i}, \{X_i\})$.
Therefore, we have $Q^*(\pi,\{X_i\}) > Q_i(\cA'',\pi,\xset)\geq Q^*(\pi_{h,-i}, \{X_i\})$.
\end{proof}
\section{Proof of Lemma~\ref{lemma:OiIsI-Part-SIR}}\label{app_lmm3}
\begin{proof}
    First, note that $\cO_i$ is IR as $\ir(\cA)$ is IR, and pruning the query tree does not increase label complexity.
    For any $i\in [k]$, suppose that there exists an hypothesis $h\in \hat H$ s.t. $\abs{H(X_i|h)} < \abs{H(X_i)}$.
    Then in the query tree of $\cA(\pi, \{X_i\})$, either all leaves are inconsistent with $h(X_{-i})$ or there exists one internal node $v$ who has exactly one subtree with all leaves inconsistent with $h(X_{-i})$.
    This node $v$ as well as the corresponding subtree are pruned in $\cO_i$ and thus the leaves in the other subtree rooted at $v$  have their depth reduced by at least $1$. Now, there exists an hypothesis $h''\in \hat H$ such that $h''(X_i)\in H(X_i|h)$ and $h''(X_i)$ is in the other subtree.
    Since $h''(X_{-i}) = h(X_{-i})$, when the underlying hypothesis is $h''$, the pruned tree given $h''(X_{-i})$ is the same as that given $ h(X_{-i})$. Hence, we have $Q_i(\cO_i,\pi,\xset, h'') \leq Q_i(\cA,\pi,\xset,h'')-1$.
    
    Then we have 
    \begin{align*}
        &Q_i(\cO_i,\pi,\xset) =  \EEs{h\sim \pi}{Q_i(\cO_i,\pi,\xset, h)}
        \\
        =& \pi(h'')Q_i(\cO_i,\pi,\xset, h'') + (1-\pi(h''))\EEs{h\sim \pi|h \neq h''}{Q_i(\cO_i,\pi,\xset, h)}\\
        \leq & \pi(h'')(Q_i(\cA,\pi,\xset, h'')-1) + (1-\pi(h''))\EEs{h\sim \pi|h\neq h''}{Q_i(\cA,\pi,\xset, h)}\\
        <& Q_i(\cA,\pi,\xset)\,,
    \end{align*}
where the last inequality holds due to $\pi(h'')>0$ since  w.l.o.g., we assumed $\pi(h)>0$ for all $h\in \hat H$ in Section~\ref{sec:model}.
\end{proof}

%%%%%%%%%%%%%%%%%%%%%%%%%%%%%%%%%%%%%%%%%%%%%%%%%%%%%%%%%%%%

\end{document}